\documentstyle[12pt]{article}
\normalbaselines
\begin{document}
\baselineskip=24pt
\bibliographystyle{unsrt}
\vbox{\vspace{6mm}}

\begin{center}
{\large\bf Density Matrix From Photon Number Tomography}
\end{center}

\bigskip
\bigskip

\begin{center}
Stefano Mancini and Paolo Tombesi
\end{center}

\begin{center}
{\it Dipartimento di Matematica e Fisica,\\
Universit\`a di Camerino, I-62032 Camerino, Italy\\
and\\
Istituto Nazionale di Fisica della Materia, Italy\\}
\end{center}

\bigskip
\bigskip

\begin{center}
Vladimir I. Man'ko \footnote{On leave from Lebedev Physical 
Institute, Moscow, 
Russia.}
\end{center}

\begin{center}
{\it Osservatorio Astronomico di Capodimonte, I-80131 Napoli,
Italy}
\end{center}

\bigskip
\bigskip
\bigskip

\begin{abstract}
We provide a simple analytic relation which connects the 
density operator 
of the radiation field
with the number probabilities. The problem of experimentally 
"sampling" 
a general matrix elements is
studied, and the deleterious effects of nonunit quantum 
efficiency in 
the detection process are
analyzed showing how they can be reduced by using the squeezing 
technique. The obtained result is particulary useful for 
intracavity field 
reconstruction states.
\end{abstract}

PACS number(s): 03.65.Bz, 42.50.Dv

\newpage

After the seminal paper of  Vogel  and Risken 
\cite{vogel} and the experimental
realization of their result by Smithey {\it et al.} \cite{raymer} it 
becames clear that the homodyne
measurement of an electromagnetic field permits the reconstruction 
of the 
Wigner  function of a
quantum state, by just varying the phase of the local oscillator, 
since 
then named optical homodyne
tomography. In Ref.
\cite{vogel} it was shown, indeed, that the rotated quadrature 
distribution may be expressed in
terms of any s-parametrized Wigner function 
\cite{cagla2}.

These papers give rise to a pletora of other papers 
\cite{tutti,antisq} where 
the limitations due to the
detectors quantum efficiency $\eta$ were also taken into account, 
and a 
fundamental limit  was
established showing the impossibility of the tomographic 
reconstruction 
for detectors quantum
efficiency less than $0.5$ \cite{darianoetal}. 
Very recently it was shown \cite{Herzog} that, in principle,
there is no such limit; however, in order to get the desired 
results for 
$\eta$ arbitrarily small, one needs to employ a rather complicate 
mathematical procedure for loss error compensation, 
which in practice make still persistent a lower limit on 
$\eta$ values because of numerical problems.
Anyway, all the above mentioned 
papers rely on homodyne
measurements, or any other phase dependent measurement process. 
While 
this work was in progress we
became aware that Wallentowitz and Vogel \cite{WVogel} 
proposed the 
s-parametrized Wigner function
reconstruction similar, in its essence, to the present 
one by using 
direct photon 
counting and contemporarely an analogous scheme was adopted by 
Banaszek and W\'odkiewicz \cite{wodkie}; however, in the 
following we 
shall discuss the relevant differences with our work.

We shall show in this
letter that the state retrieval by photon counting can be used not 
only when the 
output beam is mixed with a
reference field at a beamsplitter, as in Refs.
\cite{WVogel,wodkie}, but also in
a physical situation similar to the one proposed by Brune {\it et 
al.} \cite{haroche} for the
generation and measurement of a Schr\"odinger cat state allowing 
thus the 
possibility of
reconstructing its density matrix and, more generally, 
the possibility to 
reconstruct cavity QED field state when there is not a direct 
access to 
the field. 

We shall show that the
sampling of density matrix elements can be achieved by photon 
number measurements, for
detectors quantum efficiency greater than $0.5$, in any base but 
coordinate representation. This
$\eta \geq 0.5$ limitation, however, can be beaten if we use 
a physical technique instead of applying the 
mathematical procedure of Ref. \cite{Herzog}.

We shall call the procedure {\it Photon Number Tomography} because 
it permits the reconstruction 
of the density matrix elements, in a suitable representation, 
by just 
detecting the number of
photons at the given reference field and then scanning both 
its phase 
and its amplitude;
differently from the usual homodyne tomography where a marginal 
distribution is recorded by homodyne measurements and then 
scanning
only the phase. 

As claimed in Ref. \cite{cagla2}, it is possible to write
\begin{equation}\label{denop}
\hat\rho=\int\frac{d^2\alpha}{\pi}W(\alpha,s)\hat T(\alpha,-s)\,,
\end{equation} where the $s$-ordered wheight function $W(\alpha,s)$  
may 
be identified with the 
quasiprobability distributions $Q(\alpha)$, $W(\alpha)$ and 
$P(\alpha)$ 
when the ordering parameter
$s$ assumes the values $-1, 0, 1$ respectively; while the 
operator $\hat 
T$ represents the complex
Fourier transform of the $s$-ordered  displacement operator 
$\hat 
D(\xi,s)=\hat D(\xi)e^{s|\xi|^2/2}$,
which can also be written as \cite{cagla1}
\begin{equation}\label{T2}
\hat T(\alpha,s)=\frac{2}{1-s}\hat 
D(\alpha)\left(\frac{s+1}{s-1}\right)^{\hat a^{\dag}\hat a}
\hat D^{-1}(\alpha)\,.
\end{equation} On the other hand the weight function 
$W(\alpha,s)$ is the 
expectation value of the
operator
$\hat T(\alpha,s)$ \cite{cagla2}, i.e.
$W(\alpha,s)={\rm Tr}\{\hat\rho\hat 
T(\alpha,s)\}$,
then we obtain
\begin{equation}\label{rho}
\hat\rho=\int \frac{d^2\alpha}{\pi}\sum_{n=0}^{\infty}
\frac{2}{1-s}\left(\frac{s+1}{s-1}\right)^n
\langle n|\hat D(\alpha)\hat\rho\hat D^{-1}(\alpha)|n\rangle \hat 
T(-\alpha,-s)\,.
\end{equation} 
Thus, the weight function $W(-\alpha,s)$ is related to the ability 
of measuring the quantity $\langle n|\hat D(\alpha)\hat\rho\hat 
D^{-1}(\alpha)|n\rangle$ by scanning the whole phase space 
\cite{Royer}, 
i.e. by just varying
$\alpha$. Now, we may consider one mode of the radiation, 
whose state 
$\hat\rho$ 
one wants to
reconstruct, contained inside a cavity  and, immediately before the 
photon number measurement, a coherent
reference field is "added" \cite{haroche,Glauber63}, so that  we may 
recognize 
\begin{equation}\label{P} 
{\cal P}(n,\alpha)={\rm Tr}\{\hat 
D(\alpha)\hat\rho\hat
D^{-1}(\alpha)|n\rangle\langle n|\}
=\langle n|\hat D(\alpha)\hat\rho\hat 
D^{-1}(\alpha)|n\rangle
\end{equation} 
as the probability to detect $n$ photons after 
the injection of the
reference field $\alpha$. 
The addition process we are considering, following Ref. 
\cite{haroche},
"is quite different from 
the combination of fields produced by a beam splitter, which mixes 
together distinct modes coupled to its two ports and introduces 
vacuum 
noise  even in the absence of any classical input field". 
We are indeed
describing a much simpler field amplitude superposition mechanism, 
discussed in the Glauber's pioneering work \cite{Glauber63}.
The photon number distribution (\ref{P}) results as the 
projection of the 
field state $\hat\rho$ over a displaced number state \cite{Knight}.

The photon counting could be made either 
by means of atoms \cite{BruneWalls} as in the case of 
microwave cavity 
field or by direct detection
of the outgoing optical total field. 
Furthermore, setting 
\begin{equation}\label{K}
\hat K_s(n,\alpha)=\frac{2}{1-s}
\left(\frac{s+1}{s-1}\right)^n\hat T(-\alpha,-s)\,,
\end{equation} we have, from Eq. (\ref{rho})
\begin{equation}\label{rhosample}
\hat\rho=\sum_{n=0}^{\infty}\int\frac{d^2\alpha}{\pi}{\cal 
P}(n,\alpha)\hat K_s(n,\alpha)\,.
\end{equation} 
Thus, analogously to Ref. \cite{darianoetal}, we may
assert that a density matrix element can be experimentally 
sampled if  
there exist at least one 
value  of the parameter $s$ inside the range $[-1,1]$ for  
which the 
corresponding matrix element
of the kernel operator $\hat K_s$ is bounded. 
From Eqs. (\ref{K}) and (\ref{T2}), one immediately recognizes 
that this 
is possible for $s\in(-1,0]$,
in the number, coherent and squeezed representations and not in the 
coordinate basis. For $s=-1$, in Eq. (\ref{K}) $n=0$ only survives,
however as was shown in Ref. \cite{cagla2}, the 
operator $\hat T$ becomes singular and can be used to construct
an arbitrary density matrix when is only weighted with the well 
behavied function $Q$. It means that ${\cal P}(0,\alpha)\equiv 
Q(\alpha)$ in Eq. (\ref{rhosample}).

Let us now
consider the more realistic case of nonunit quantum efficiency 
$\eta$. 
Accordingly 
to Ref. \cite{SL} we have
\begin{equation}\label{PofPmu} 
{\cal 
P}(n,\alpha)=\eta^{-n}\sum_{k=0}^{\infty}{\cal
P}_{\eta}(n+k,\alpha)
\left(\begin{array}{cc}n+k\\ n\end{array}\right)
\left(\frac{1-\eta}{\eta}\right)^k(-1)^k\,.
\end{equation}  
where ${\cal P}_{\eta}$ represents the same of (\ref{P}) 
but in presence of
efficiency $\eta$, i.e.
\begin{equation}\label{Peff} {\cal P}_{\eta}(n,\alpha)={\rm Tr}
\left\{\hat D(\alpha)\hat\rho\ D^{-1}(\alpha) 
:e^{-\eta\hat a^{\dag}\hat a}
\frac{(\eta\hat a^{\dag}\hat a)^n}{n!}:\right\}\,.
\end{equation} 
 If we substitute Eq. (\ref{PofPmu}) into Eq. 
(\ref{rhosample}), we obtain
\begin{equation}\label{rhosamplemu}
\hat\rho=\sum_{m=0}^{\infty}\int\frac{d^2\alpha}{\pi}{\cal 
P}_{\eta}(m,\alpha)
\left[
\sum_{l=0}^{m}\eta^{-(m-l)}\left(\frac{\eta-1}{\eta}\right)^l
\left(\begin{array}{cc}m\\m-l\end{array}\right)
\hat K_s(m-l,\alpha)\right]\,,
\end{equation} where the quantity inside the square brackets can be 
considered as a modified kernel
\begin{equation}\label{Kmu}
\hat K_{s,\eta} 
(m,\alpha)=\frac{2}{1-s}\left(\frac{\eta s-\eta+2}{\eta s-\eta}
\right)^m\hat 
T(-\alpha,-s)\,.
\end{equation} 
Again, this kernel results bounded only in the number, 
coherent and squeezed
representations,  but in that case the $s$ values for which 
that occours 
are 
determined by the quantum
efficiency; in particular we have from Eq. (\ref{Kmu}) 
$s\in(-1,-(1-\eta)/\eta]$,  
so that $\eta$ should be greater than $0.5$, similar to Ref.
\cite{darianoetal}.

It is also interesting to note as was stressed in Ref. \cite{WVogel},
that 
if one chooses 
$s=1-2/\eta$ in Eq. (\ref{Kmu}), only the term
with $m=0$ survives in Eq. (\ref{rhosamplemu}), and this means that 
the desired density matrix
elements are simply related to the zero-count probability which 
gives 
only the $Q$-function. 
Unfortunately, only
quantum efficiency $\eta=1$ (i.e. $s=-1$) allows to get the 
$Q$-function
in this 
simpler manner, otherwise, for $\eta<1$, $s$ takes forbidden 
values (i.e.
becomes smaller than $-1$). 

Let us now show how the above limit on detectors efficiency can be 
beaten. 
In case one squeezes the radiation inside the cavity 
immediately before
the injection of the reference field, the superposition mechanism
\cite{Glauber63} can be read as 
\begin{equation}\label{sqobs}
\hat D(\alpha)\hat S(\zeta)\hat\rho\hat S^{-1}(\zeta)
\hat D^{-1}(\alpha)=
\hat D(\alpha)\hat{\tilde\rho}
\hat D^{-1}(\alpha)\,,
\end{equation} where $\hat S(\zeta)$ is the well known 
squeeze operator 
\cite{Yuen} with the
squeezing parameter $\zeta=|\zeta|e^{i\varphi}$. Of course one 
has to be 
sure that the squeezing process is like a kick during which 
the natural 
evolution of the system is negligible; only under this assumption 
Eq. (\ref{sqobs}) holds. It could be physically realized, for 
example, by using a 
$\delta$-kicking of frequency of the cavity mode \cite{olga}.

Since the weight function
$W(\alpha,s)$ can be written as the complex Fourier transform of the 
characteristic function, 
let us examine the relation between the operator of Eq. 
(\ref{sqobs}) and 
$\hat D(\alpha)\hat\rho
\hat D^{-1}(\alpha)$
in terms of characteristic functions. By definition we have
\begin{equation}
\tilde\chi(\xi,s)={\rm Tr}\{\hat{\tilde\rho} \hat D(\xi,s)\}
={\rm Tr}\{\hat\rho
\hat S^{-1}(\zeta)e^{\xi\hat a^{\dag}}e^{-\xi^*\hat a}
\hat S(\zeta)\}e^{-|\xi|^2/2}e^{s|\xi|^2/2}\,.
\end{equation} Remembering how the squeeze operator acts on  
the annihilation (creation) operator \cite{Yuen},
we obtain
\begin{equation}
\tilde\chi(\xi,s)={\rm Tr}\{\hat\rho\hat D(\xi\mu^*+
\xi^*\nu)\}e^{s|\xi|^2/2}
=\chi(\xi\mu^*+\xi^*\nu)e^{s|\xi|^2/2}\,,
\end{equation} 
with $\mu=\cosh|\zeta|$ and $\nu=e^{i\varphi}\sinh|\zeta|$. 
Writing now the variable $\xi$ in polar coordinate 
$\xi=|\xi|e^{i\phi}$ and locking the squeezing parameter phase 
with that 
of the reference
field, i.e. $\varphi=2\phi$, we get 
$\xi\mu^*+\xi^*\nu=\xi
(\cosh|\zeta|+\sinh|\zeta|)=\xi e^{|\zeta|}=\xi\Delta$, 
such that
$\chi(\xi,s)=\tilde\chi(\xi/\Delta,s\Delta^2)$ and 
$W(\alpha,s)
=\Delta^2\tilde W(\Delta\alpha,
s\Delta^2)$.
Hence, from Eqs. (\ref{denop}), (\ref{rho}), (\ref{PofPmu}) 
and the last expression, we get
\begin{equation}\label{rhosamplesq}
\hat\rho=\sum_{n=0}^{\infty}\int\frac{d^2\alpha}{\pi}
{\cal P}_{\eta}(n,\alpha){\hat{\cal
K}}_{s,\eta}(n,\alpha/\Delta)\,,
\end{equation}
where 
\begin{equation}\label{Ksq}
{\hat{\cal K}}_{s,\eta}(n,\frac{\alpha}{\Delta})=
\frac{2}{1-s\Delta^2}\left(\frac{\eta s\Delta^2-\eta+2}{\eta 
s\Delta^2-\eta}
\right)^n\hat 
T(-\frac{\alpha}{\Delta},-s)\,,
\end{equation}
from which one immediately recognizes that the ordering 
parameter $s$ in the power term, is now scaled by
the factor $\Delta^2$ with respect to the one in Eq. (\ref{Kmu}), 
thus it ranges in the interval
$s\in(-1,-(1-\eta)/\eta\Delta^2]$,
allowing values of quantum efficiency lower than 0.5.
It is then possible to define an effective 
quantum
efficiency
$\tilde\eta=\Delta^2/[\Delta^2+(1-\eta)/\eta]$
which could be close to unit for sufficientely large squeezing 
independently of 
the real quantum efficiency.

For an easier implementation of this scheme, 
we could choose in Eq.
(\ref{Ksq}) $s=\Delta^{-2}(1-2/\eta)$, in order to eliminate 
the sum over 
$n$ in Eq.
(\ref{rhosamplesq}), directly relating the density matrix 
elements in any 
allowed basis to 
the zero-count probability; but
in this case, due to the presence of the factor $\Delta^{-2}$, 
the allowed values of $s$
are not $s=-1$ as in Ref. \cite{WVogel}, which only permits the 
reconstruction of the more smoothed $Q$-function, rather 
the value of $s$ 
can approach zero from below depending on the value of the 
squeezing parameter, independently of $\eta$.

Thus, by squeezing preventively the field in a 
cavity allows one to detect the state of radiation inside 
it even for 
low efficiency detection.
The use of squeezing technique to reduce the effect of 
nonunit quantum 
efficiency was already proposed in Ref. \cite{antisq}, 
but we use it in a 
different way. In Ref. \cite{antisq}, 
indeed, antisqueezing preamplification of the signal at 
the beam splitter was proposed rather than the  
antisqueezed signal 
inside the cavity.
We would stress the fact that our Photon Number 
Tomography scheme 
becomes especially useful in
the intracavity optical tomography where other similar schemes 
\cite{WVogel,wodkie} are not applicable; in fact it can be 
adopted in a situation
in which the photon number is measured indirectly 
using a "sequence" of atoms
passing through the cavity
\cite{BruneWalls}, with the quantum efficiency, determined 
only by the duration of the measurement process 
(i.e. the length of the 
"sequence"), that could be very high \cite{haroche}. 
In that case, the 
scheme has also the advantage of being QND. 
Finally our scheme results suitable for cavity QED 
characterization, like Ref. \cite{meysc},
 allowing the reconstruction of 
nonclassical states as well,
 which are extremely sensitive to losses and then 
their detection seem prohibitive  by means of an 
outgoing field as in 
Refs. \cite{WVogel,wodkie}.

At last, our approach could be considered an extension of the 
known mathematical
principle of tomography. In fact given a density operator $\hat\rho$ 
and a group element $\cal G$,
one can create different types of tomography if, 
by knowing the matrix 
elements 
$\langle x|{\cal G}\hat\rho{\cal G}^{-1}|x\rangle$ 
from measurements, 
is able to invert the formula
expressing the density operator in terms of the above distribution  
(the $x$ may denote either
continous or discrete eigenvalues). For the inversion procedure 
one can use the properties of
summation or integration over group parameters $\cal G$.
The known tomographies are just given by this construction; 
for Radon transform or homodyne tomography, $\cal G$ is the
rotation group, for symplectic tomography \cite{MMT} it is 
the symplectic group, but in principle
one can use other groups. The only problem is mathematical 
one to make the inversion and/or
physical one to realize the transformation $\cal G$ in laboratory.

\end{document}